\documentclass[preprint]{aastex}
\newcommand{\bm}[1]{\mbox{\bf #1}}
\shorttitle{Nonlinear Stochastic Biasing of Galaxies}
\shortauthors{Yoshikawa, Taruya, Jing \& Suto}
\begin{document}
\title{NONLINEAR STOCHASTIC BIASING OF GALAXIES AND DARK HALOS \\ IN
COSMOLOGICAL HYDRODYNAMIC SIMULATIONS}
\author{Kohji Yoshikawa}
\affil{Department of Astronomy, Kyoto University, Kyoto 606-8502,
Japan.}
\email{kohji@kusastro.kyoto-u.ac.jp}
\author{Atsushi Taruya}
\affil{Department of Physics, School of Science, University of Tokyo,
Tokyo 113-0033, Japan.}
\email{ataruya@utap.phys.s.u-tokyo.ac.jp}
\author{Y.P. Jing} 
\affil{Shanghai Astronomical Observatory, the Partner Group of MPI f\"ur
Astrophysik, \\Nandan Road 80, 200030 Shanghai, China.
\\ and \\ National Astronomical
Observatories, Chinese Academy of Sciences, 100012 Beijing, China.}
\email{ypjing@center.shao.ac.cn}
\and
\author{Yasushi Suto} 
\affil{Department of Physics and Research Center for the Early Universe
(RESCEU)\\ School of Science, University of Tokyo, Tokyo 113-0033,
Japan.}
\email{suto@phys.s.u-tokyo.ac.jp}

\received{2000 December 12}
\accepted{2001 ???}

\begin{abstract}
  We perform an extensive analysis of nonlinear and stochastic biasing
of galaxies and dark halos in spatially flat low-density CDM universe
($\Omega_0=0.3, \lambda_0=0.7, h=0.7$, and $\sigma_8=1$) using
cosmological hydrodynamic simulations. We identify galaxies by linking
cold and dense gas particles which satisfy the Jeans criterion.  We
compare their biasing properties with the predictions of an analytic
halo biasing model.  Dark halos in our simulations exhibit reasonable
agreement with the predictions only on scales larger than $\sim
10h^{-1}$Mpc, and on smaller scales the volume exclusion effect of halos
due to their finite size becomes substantial.  Interestingly the biasing
properties of galaxies are better described by {\it extrapolating} the
halo biasing model predictions.

The clustering amplitudes of galaxies are almost independent of the
redshift between $z=0$ and 3 as reported in previous simulations. This
in turn leads to a rapidly evolving biasing factor; we find that $b_{\rm
cov}\simeq 1$ at redshift $z\simeq 0$ to $b_{\rm cov}\simeq 3-4$ at
$z=3$, where $b_{\rm cov}$ is a biasing parameter defined from the
linear regression of galaxy and dark matter density fields. Those values
are consistent with the observed clustering of Lyman-break galaxies.

We also find the clear dependence of galaxy biasing on their formation
epoch; the distribution of old populations of galaxies tightly
correlates with the underlying mass density field, while that of young
populations is slightly more stochastic and anti-biased relative to dark
matter. The amplitude of two-point correlation function of old
populations is about 3 times larger than that of the young
populations. Furthermore, the old population of galaxies reside within
massive dark halos while the young galaxies are preferentially formed in
smaller dark halos.  Assuming that the observed early and late-type
galaxies correspond to the simulated old and young populations of
galaxies, respectively, all of these segregations of galaxies are
consistent with observational ones for the early and late-type of
galaxies such as the morphology--density relation of galaxies.
\end{abstract}

\keywords{galaxies: clustering -- galaxies: formation -- galaxies: halos
-- dark matter -- cosmology: large-scale structure of universe --
methods: numerical}

\section{INTRODUCTION}

Ongoing galaxy redshift surveys, such as the Sloan Digital Sky Survey
(SDSS) and the Two-Degree Field (2dF), aim at revealing the the
large-scale structure of the universe with unprecedented precision. The
gravitational instability is the main key process of the {\it dark
matter clustering}, and this is now well understood from numerical
simulations and several empirical theoretical models \citep{Davis1985,
Hamilton1991, Suto1993, Mo1996, Navarro1997}. In fact once the
underlying cosmological models are specified, the two-point correlation
functions of dark matter, which are the most conventional and widely
used statistics describing the large-scale structure, can be fairly
accurately predicted even with the redshift distortion and light-cone
effects \citep{Peacock1996, Suto1999, Suto2000, Hamana2001}.

On the other hand, it is widely believed that the distribution of
galaxies is somewhat biased with respect to the underlying dark matter.
For instance, Lyman-break galaxies at redshift $z \approx 3$
\citep{Steidel1998, Adelberger1998, Giavalisco1998} exhibit strong
clustering and the galaxy biasing with respect to dark matter is
time-dependent.  Also the galaxy clustering is dependent on the galaxy
morphology and environment \citep{Dressler1980, Postman1984,
Loveday1995, Hermit1996, Dressler1997, Tegmark1999} indicating the
galaxy biasing is sensitive to many physical processes and thus {\it
stochastic}.  Clearly the relation between galaxy and dark matter
clustering is far from simple and not yet fully understood either
observationally or theoretically.  This is the primary difficulty in
properly interpreting the observational data of the upcoming large-scale
redshift surveys.

So far several models of galaxy biasing have been proposed adopting
simplifying assumptions; \citet{Fry1996} and \citet{Tegmark1998} discuss
the evolution of biasing assuming that the number of galaxies does not
change. \citet{Mo1996} present a model for the nonlinear biasing of
virialized dark halos using the extended Press--Schechter formalism
\citep{Bond1991}. \citet{Jing1998} tested and improved the formula for
the biasing of halo correlation functions originally proposed by
\citet{Mo1996} using high-resolution $N$-body simulations.
\citet{Dekel1999} develop a fundamental framework to quantify the
nonlinearity and stochasticity in galaxy biasing.  Their formulation was
subsequently applied to several numerical simulations
\citep{Blanton1999, Blanton2000, Somerville2001}. The biasing of dark
halos is also investigated by \citet{Kravtsov1999} using high resolution
$N$-body simulations. Note that their definition of dark {\it halos} is
different from the conventional one used in the Press--Schechter
formalism but rather close to {\it dark matter cores} (DM cores) in our
analysis below.  Recently, Taruya \& Suto (2000; TS hereafter) proposed
a first physical and analytical model for nonlinear and stochastic halo
biasing combining the biasing model of \citet{Mo1996} and the formation
epoch distribution \citep{Kitayama1996}.

More realistic approaches to galaxy biasing employ the state-of-the-art
numerical simulations including the mesh-based hydrodynamical
simulations \citep{Blanton1999, Blanton2000, Cen2000}, and $N$-body
simulations combined with semi-analytic modeling of galaxy formation
\citep{Benson2000, Somerville2001}.  In what follows, we use the
cosmological smoothed particle hydrodynamic (SPH) simulations
\citep{Yoshikawa2000} of cold dark matter (CDM) universe to examine the
galaxy biasing. In particular, we focus on the comparison of the biasing
characteristics of simulated objects (galaxies and dark halos) with the
halo biasing model of TS. In addition, we investigate dependence of
galaxy biasing properties on their formation history as an origin of
galaxy morphology. Our simulation directly follows hydrodynamical and
radiative processes to simulate galaxy formation, while the evolution of
galaxies is not so properly modeled as those combined with a
semi-analytic method of galaxy formation \citep{Somerville2001}.  Due to
the Lagrangian nature of the SPH technique, the spatial resolution of
our simulations is better than those in the mesh-hydro simulations by
\citet{Blanton1999, Blanton2000} and we can resolve galaxies as distinct
and isolated objects, while their treatment of the thermal process and
the metal enrichment is more realistic.  Thus our method is
complementary to those previous investigations with different
approaches.

The rest of the paper is organized as follows. In \S 2, we describe the
detail of our numerical simulation and the procedures to identify
galaxies, dark matter cores (DM cores), and dark halos. In \S 3 we
present a brief summary of the biasing description following TS, and
compare several properties of the biasing in the one-point statistics of
galaxies and dark halos. Then we discuss the biasing in terms of their
two-point correlation functions. Section 4 examines the dependence of
galaxy biasing on their formation history. Finally, we summarize our
major findings in \S 5.

\section{METHODOLOGY}
\subsection{Cosmological SPH simulation}

Our numerical simulation code is a hybrid of
Particle--Particle--Particle--Mesh (P$^3$M) $N$-body Poisson solver
\citep{Hockney1981} and smoothed particle hydrodynamics (SPH) algorithm
\citep{Yoshikawa2000}. The simulation presented in this paper adopts
$N_{\rm DM}=128^3$ dark matter particles and the same number of gas
particles for SPH. We use the spline (S2) functional form for
gravitational softening \citep{Hockney1981} and the softening length is
set to $\epsilon_{\rm grav}=L_{\rm box}/(10N_{\rm DM}^{1/3})$ and kept
constant in comoving coordinates, where $L_{\rm box}$ is the comoving
size of the simulation box. We set the minimum of SPH smoothing length
to $h_{\rm min}=\epsilon_{\rm grav}/4$ and adopt the ideal gas equation
of state with an adiabatic index $\gamma=5/3$. The effect of radiative
cooling is included adopting the metallicity of [Fe/H]$=-0.5$. We use
the cooling rate described in
\citet{Sutherland1993}. \citet{Thacker2000} reported that artificial
over-cooling occurs under the presence of radiative cooling in SPH
simulations due to overestimate of hot gas density in the vicinity of
cooled gas clumps due to the smoothing scheme of SPH algorithm.  In
order to avoid this numerical artifact, we implement a modification of
SPH algorithm, ``cold gas decoupling'', following \citet{Pearce1999}.
The detail of this prescription is presented in the next subsection.

We consider a spatially-flat low-density CDM (LCDM) universe with
$\Omega_0=0.3$, $\lambda_0=0.7$, $\sigma_8=1.0$ and $h=0.7$, where
$\Omega_0$ is the mean mass density parameter, $\lambda_0$ the
dimensionless cosmological constant, $\sigma_8$ the rms density
fluctuation on a scale of $8h^{-1}$ Mpc and $h$ the Hubble constant in
units of 100 km$\cdot$s$^{-1}\cdot$Mpc$^{-1}$. This particular model
satisfies both the {\it COBE} normalization \citep{Bunn1997} and the
abundance of X-ray clusters of galaxies \citep{Kitayama1997}. We assume
the mean baryon mass density parameter to be $\Omega_{\rm
b}=0.015h^{-2}$ \citep{Copi1995}. The simulation is carried out in a
periodic cube of ($75h^{-1}$Mpc)$^3$, with the gas and dark matter mass
per particle being $2.4\times 10^9 M_{\odot}$ and $2.2\times 10^{10}
M_{\odot}$, respectively. The initial condition is created at $z=25$
using the COSMICS package \citep{Bertschinger1995}, which is evolved up
to $z=0$.

\subsection{Cold gas decoupling and identification of galaxies}
\label{sec:galdef}

In order to avoid the numerical over-cooling of gas particles mentioned
above, we decouple cold gas particles which satisfy the
following Jeans condition \citep{Yoshikawa2000}:
\begin{equation}
 \label{eq:decouple}
 h_{\rm\scriptscriptstyle SPH} >\frac{c_s}{\sqrt{\pi G\rho_{\rm gas}}},
\end{equation}
where $h_{\rm\scriptscriptstyle SPH}$ is the smoothing length of gas
particles, $c_s$ the sound speed, $G$ the gravitational constant and
$\rho_{\rm gas}$ the gas density of gas particles. Except for the fact
that these cold gas particles are ignored in computing the gas density
of hot gas particles, all the other SPH interactions are left
unchanged. This decoupling scheme is a phenomenological treatment of
multi-phase gas dynamics, and should be interpreted as an approximate
prescription of galaxy formation.

\begin{figure}[tbp]
 \leavevmode
 \begin{center}
  \includegraphics[keepaspectratio,width=12cm]{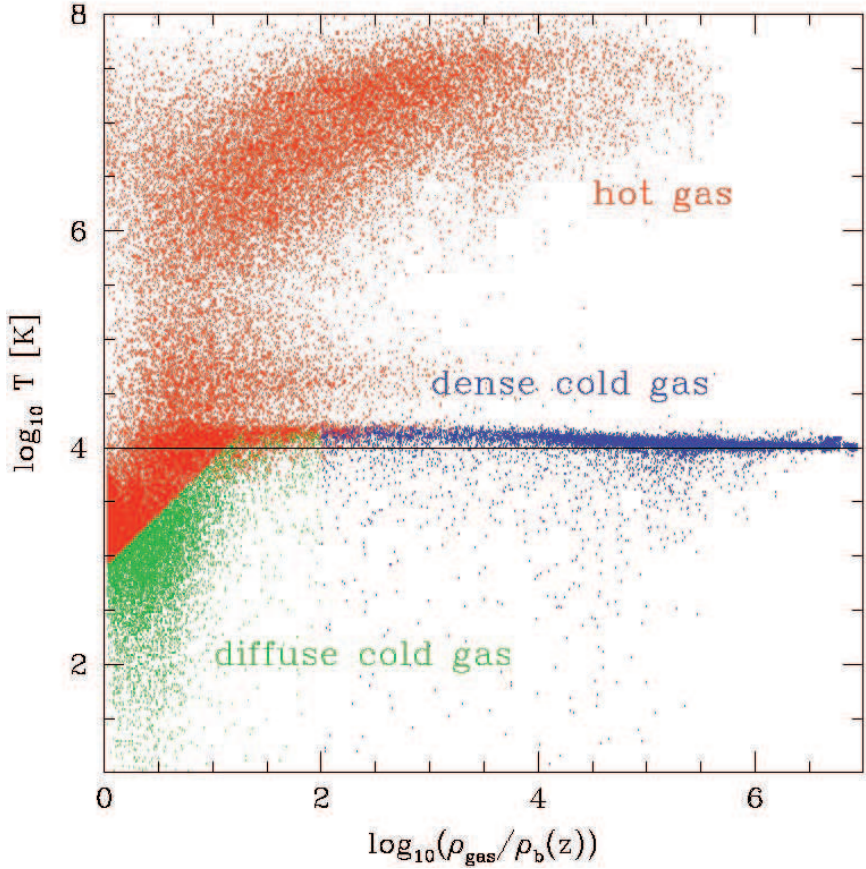}
  \figcaption{Scatter plot of gas particles at $z=0$ in $\log\rho_{\rm
  gas}$--$\log T$ plane.  The cold and dense gas particles which satisfy
  the criterion (\ref{eq:decouple}) and $\rho_{\rm
  gas}>10^2\,\bar{\rho}_{\rm b}(z)$ are indicated by blue points, the
  diffuse cold gas with $\rho_{\rm gas}>10^2\,\bar{\rho}_{\rm b}(z)$ are
  by green points and the others are by red points.
  \label{fig:scatterplot}}
 \end{center}
\end{figure}

Galaxies in our simulations are identified as clumps of cold and dense
gas particles which satisfy the criterion (\ref{eq:decouple}) and
\begin{equation}
 \label{eq:densthresh}
  \rho_{\rm gas}>10^2\,\bar{\rho}_{\rm b}(z),
\end{equation}
where $\bar{\rho}_{\rm b}(z)$ is the mean baryon density at redshift
$z$.  Figure~\ref{fig:scatterplot} shows the scatter plots of gas
particles in density -- temperature plane.  The blue points indicate the
cold and dense gas particles satisfying the criteria (\ref{eq:decouple})
and (\ref{eq:densthresh}), the diffuse cold gas particles which satisfy
(\ref{eq:decouple}) and $\rho_{\rm gas}<10^2\,\bar{\rho}_{\rm b}(z)$ are
plotted in green, and the other hot gas particles are represented in
red. This indicates that the above criteria for the galaxy particles
properly segregate the cold and dense gas particles. We group these
particles using friend-of-friend (FOF) algorithm \citep{Davis1985} with
linking length $b_{\rm g}=0.0164(1+z)\,\bar{l}$ and identify the
resulting clumps as ``galaxies'', where $\bar{l}=L_{\rm box}/N_{\rm
DM}^{1/3}$ is the comoving mean particle separation. The proper choice
of the linking length is not clear and we simply adopt the value of
\citet{Pearce1999} here.  In this paper, we only consider galaxies with
mass greater than $M_{\rm g,min}=10^{11} M_{\odot}$, which is
equivalently 40 times of each gas particle mass and close to a nominal
mass resolution of baryonic matter\footnote{SPH gas density is smoothed
over about 30 nearest neighbor gas particles.}. As noted in
\S~\ref{subsec:DMcore}, the mass functions of simulated galaxies are
roughly consistent with those from semi-analytic modeling of galaxy
formation, which justifies our galaxy criteria empirically to some
extent. We show the number of galaxies identified in our simulation and
the adopted linking length in Table~\ref{tab:objects}.

\subsection{Identification of dark halos and dark matter cores}
\label{subsec:DMcore}

The FOF algorithm is also applied in identifying dark halos. The linking
length $b_{\rm h}$ for dark halos is set to satisfy the equation
\begin{equation}
 \frac{\Delta_c(z)}{18\pi^2} = \left(\frac{b_{\rm h}}{0.2\,\bar{l}}\right)^{-3},
\end{equation}
where $\Delta_c(z)$ is the mean over-density of spherically virialized
objects formed at redshift $z$. We compute $\Delta_c(z)$ at each
redshift using a fitting formula by \citet{Kitayama1996}. At $z=0$, for
instance, $\Delta_c=335$ and $b_{\rm h}=0.164\,\bar{l}$. 

We also identify the surviving high-density substructures in dark halos,
which we call DM cores. Identification of substructures in dark halos is
a technically challenging problem and several objective methods have
been proposed so far \citep{Gelb1994, Eisenstein1998, Klypin1999,
Springel2000}. In order to identify DM cores in our simulation, we adopt
the hierarchical FOF (HFOF) method \citep{Gottlober1999}. In HFOF
method, we apply the conventional FOF method with a set of different
linking length $b_{\rm c}$: $b_{\rm c}=l_{\rm max}/4$, $l_{\rm max}/2$,
and $l_{\rm max}$, where $l_{\rm max}$ is the maximum linking length.
For each linking length, gravitationally bound groups with more than 20
particles are identified as DM cores. The maximum linking length is set
to $l_{\rm max}=0.05\bar{l}$.

In this paper, we consider the dark halos with their mass greater than
$10^{12} M_{\odot}$ ($\simeq \Omega_0/\Omega_{\rm b}\times M_{\rm
g,min}$) and DM cores with more than 20 dark matter particles
(equivalently $4.3\times10^{11} M_{\odot}$). In Table~\ref{tab:objects},
the number of identified objects and adopted linking length are also
shown. Figures~\ref{fig:LCDM_Z00} and \ref{fig:LCDM_Z20} show the
distribution of dark matter particles, gas particles, dark halos and
galaxies at $z=0$ and $z=2$. At $z=0$ galaxies are more strongly
clustered than dark halos, while at $z=2$ those two objects show similar
spatial distribution.

Figures~\ref{fig:region_1} and \ref{fig:region_2} show close-up
snapshots of the most massive cluster at $z=0$ with mass $M\simeq
8\times 10^{14}M_{\odot}$ and a relatively poor cluster with $M\simeq
10^{14}M_{\odot}$, respectively. In each figure, upper panels depict the
distribution of dark matter and gas particles, and the distributions of
DM cores and dense cold gas particles which satisfy the condition
(\ref{eq:decouple}) and (\ref{eq:densthresh}) are shown in lower
panels. Circles in lower panels indicate the positions of galaxies
identified in our simulation. We can see that for the richer cluster,
the distribution of DM cores is relatively in good agreement with that
of galaxies except for the cluster center, where the tidal radius is
much shorter than our numerical resolution. On the other hand, galaxies
or cold gas clumps in the smaller cluster are not necessarily hosted by
DM cores. This is probably because DM cores in our simulation
significantly suffer from the artificial overmerging, which is severer
for poorer dark halos due to small number of particles, while galaxies
represented by dissipative gas particles are less affected by this
overmerging.  This is why DM cores at higher redshift are much less
abundant than galaxies and dark halos (see Table~\ref{tab:objects}).
This problem is intrinsically related to the question of whether
substructures within dark halos identified in high-resolution $N$-body
simulations \citep{Klypin1999,Colin1999} really correspond to the real
galaxies. Unfortunately the resolution of our current simulations is not
sufficiently good to answer this issue in a reliably manner, but we hope
to revisit this with another SPH run with $N=256^3$ particles
(Yoshikawa, Jing \& Suto, in preparation).

\begin{figure}[tbp]
 \leavevmode
 \begin{center}
  \includegraphics[keepaspectratio,width=15cm]{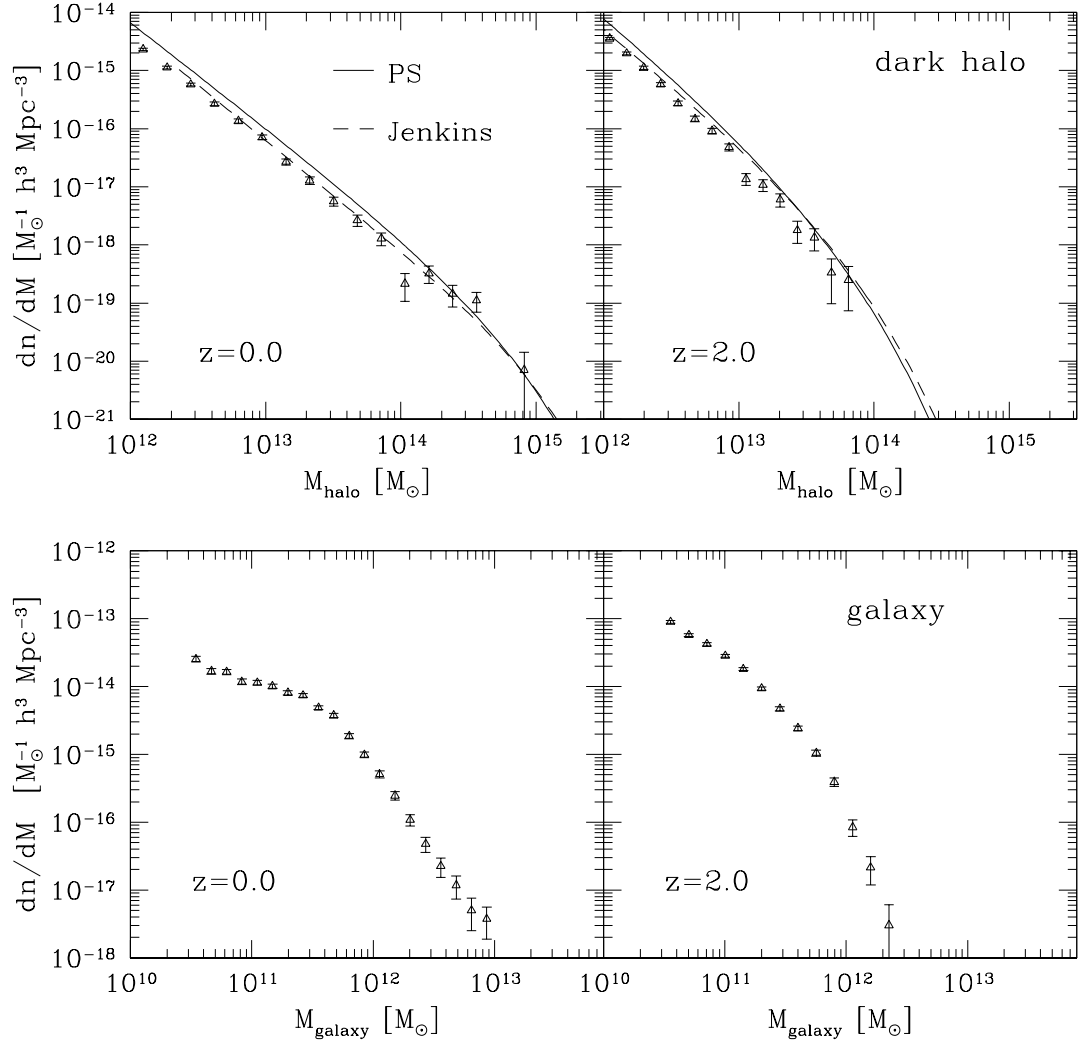}
  \figcaption{Mass functions of dark halos and galaxies at redshift
  $z=0$ and $z=2$. Solid lines in upper panels are theoretical
  predictions of Press--Schechter mass function and dashed lines are
  fitting formula by \citet{Jenkins2001}.\label{fig:massfunction}}
 \end{center}
\end{figure}

Figure~\ref{fig:massfunction} shows mass function of dark halos and
galaxies at $z=0$ and $2$.  We find that the mass function of simulated
dark halos ({\it upper panels}) agrees better to the fitting formula of
\citet{Jenkins2001} ({\it dashed lines}) than that of \citet{Press1974}
({\it solid lines}).  Galaxy mass functions in our simulations ({\it
lower panels}) are roughly consistent with those from other SPH
simulations and semi-analytic models \citep{Benson2001}, but slightly
less abundant at $M_{\rm galaxy}\lesssim 10^{11} M_{\odot}$ due to
limited mass resolution.

\section{BIASING PROPERTIES OF GALAXIES AND DARK HALOS}

The most natural form of galaxy biasing is the relation between
over-density fields of galaxies $\delta_{\rm g}$ and dark matter
$\delta_{\rm m}$. In this section, we compute the density fields of
galaxies and dark halos from our simulation, and study their
statistical properties and evolution.

\subsection{Formulation and computation of biasing parameters}

A biasing scheme relates the density field of dark matter with those of
galaxies and dark halos, which are defined for a given smoothing scale
$R_s$ as
\begin{eqnarray}
 \delta_{\rm m}({\bm x},R_s) = \frac{\rho({\bm x},R_s)}{\bar{\rho}}-1, \\
 \label{eq:galdens}
 \delta_{\rm g}({\bm x},R_s) 
= \frac{n_{\rm g}({\bm x},R_s)}{\bar{n}_{\rm g}}-1, \\
 \label{eq:halodens}
 \delta_{\rm h}({\bm x},R_s) 
= \frac{n_{\rm h}({\bm x},R_s)}{\bar{n}_{\rm h}}-1,
\end{eqnarray}
where $\rho({\bm x},R_s)$, $n_{\rm g}({\bm x},R_s)$, and $n_{\rm h}({\bm
 x},R_s)$ denote the mass density, and galaxy and halo number densities
 smoothed over the top-hat window radius $R_s$, and the over-bar
 $(\,\bar{\,}\,)$ indicates the mean over the entire universe. We
 briefly summarize several parameters describing the nonlinear
 stochastic nature of biasing introduced by TS for later convenience.

The joint probability distribution function (PDF), $P(\delta_{\rm m},
\delta_i)$, characterizes the statistical properties of $\delta_{\rm m}$
and $\delta_i$, where the subscript $i$ indicates two different objects;
``g'' for galaxies and ``h'' for dark halos. By definition, $\delta_{\rm
m}$ and $\delta_i$ have zero mean and their variances are related to
the joint PDF as
\begin{equation}
 \label{eq:begin}
 \sigma_{\rm m}^2 = \langle\delta_{\rm m}^2\rangle 
= \int\int P(\delta_{\rm m},\delta_i)\delta_{\rm m}^2\,
  d\delta_{\rm m}\,d\delta_i
\end{equation}
and
\begin{equation}
 \sigma_i^2 = \langle\delta_i^2\rangle 
= \int\int P(\delta_{\rm m},\delta_i)\delta_i^2\,
   d\delta_{\rm m}\,d\delta_i ,
\end{equation}
where the bracket $\langle\cdots\rangle$ denotes the joint average over
$\delta_i$ and $\delta_{\rm m}$.  The statistical relation between
$\delta_i$ and $\delta_{\rm m}$ is described by the conditional PDF, 
$P(\delta_i|\delta_{\rm m})$. The conditional mean of $\delta_i$,
$\bar{\delta_i}(\delta_{\rm m})$, for a given $\delta_{\rm m}$ is then
calculated from
\begin{equation}
 \bar{\delta_i}(\delta_{\rm m}) 
= \int\delta_i\,P(\delta_i|\delta_{\rm m})d\delta_i ~ ,
\end{equation}
yielding the following biasing parameter (TS):
\begin{equation}
 b_{{\rm cov},i} \equiv 
\frac{\langle \bar{\delta_i}(\delta_{\rm m})\delta_{\rm m}\rangle}
{\sigma_{\rm m}^2} 
= \frac{\langle\delta_i\delta_{\rm m}\rangle}{\sigma_{\rm m}^2} .
\end{equation}
The nonlinearity of the biasing is quantified by
\begin{equation}
 \epsilon^2_{{\rm nl},i} \equiv 
\frac{\langle\delta_{\rm m}^2\rangle\langle\bar{\delta}_i^2\rangle}
{\langle \bar{\delta}_i\delta_{\rm m}\rangle^2}-1 
= \frac{\sigma_{\rm m}^2\langle\bar{\delta}_i^2\rangle}
{\langle\delta_i\delta_{\rm m}\rangle^2}-1 ,
\end{equation}
which vanishes only when the biasing is linear (i.e., the ratio
$\delta_i/\delta_{\rm m}$ is independent of $\delta_{\rm m}$) and is
positive otherwise. Similarly the stochasticity of the biasing is
characterized by
\begin{equation}
 \epsilon^2_{{\rm scatt},i} \equiv 
\frac{\langle\delta_{\rm m}^2\rangle\langle(\delta_i-\bar{\delta}_i)^2\rangle}
{\langle\bar{\delta}_i\delta_{\rm m}\rangle^2}
= \frac{\sigma_{\rm m}^2[\sigma_i^2-\langle\bar{\delta}_i^2\rangle]}
{\langle\delta_i\delta_{\rm m}\rangle^2} .
\end{equation}
This parameter vanishes for the deterministic bias where $\delta_i =
\bar{\delta}_i(\delta_{\rm m})$.  In terms of the above biasing
parameters, a somewhat more conventional biasing coefficient $b_{{\rm
var},i}\equiv\sigma_i/\sigma_{\rm m}$ is written as
\begin{equation}
\label{eq:bvar}
 b_{{\rm var},i} = b_{{\rm cov},i}
\,(1+\epsilon^2_{{\rm nl},i}+\epsilon^2_{{\rm scatt},i})^{1/2}.
\end{equation}
Finally the correlation coefficient $r_{{\rm corr},i}$ \citep{Dekel1999}
is given by
\begin{equation}
 \label{eq:end}
  r_{{\rm corr},i} \equiv \frac{\langle\delta_i\delta_{\rm m}\rangle}
{\sigma_i\sigma_{\rm m}} 
= \frac{1}{\sqrt{1+\epsilon^2_{{\rm scatt},i}+\epsilon^2_{{\rm nl},i}}} .
\end{equation}

We compute the biasing parameters $b_{{\rm cov},i}$, $b_{{\rm var},i}$,
$\epsilon_{{\rm nl},i}$, $\epsilon_{{\rm scatt},i}$ and $r_{{\rm
corr},i}$ each for dark halos and galaxies with smoothing scales
$R=4h^{-1}$Mpc, $8h^{-1}$Mpc and $12h^{-1}$Mpc. We obtain many pairs of
the values $(\delta_i({\bm x},R_s),\delta_{\rm m}({\bm x},R_s))$ for
randomly selected points ${\bm x}$ in the simulation volume and evaluate
the biasing parameters using equations (\ref{eq:begin}) --
(\ref{eq:end}) by replacing the joint averages $\langle\cdots\rangle$
with averages over all selected points. The number of randomly selected
points is 1000 for the top-hat smoothing scale $R_s=12h^{-1}$Mpc, 5000
for $R_s=8h^{-1}$Mpc and 30000 for $R_s=4h^{-1}$Mpc. Since our
simulation volume is $75h^{-1}$Mpc per side, most of the selected
sampling points are not fully independent.  Nevertheless we decided to
make over-sampling in evaluating the mean and the variance of the
density fields. Thus our quoted error-bars below may rather correspond
to those in a bootstrap resampling method.

\subsection{Comparison of biasing of galaxies and dark halos}\label{subsec:bias}

Figure~\ref{fig:sm412} shows the joint distribution of $\delta_{\rm h}$
and $\delta_{\rm g}$ with mass density field $\delta_{\rm m}$ at
redshift $z=0$, 1 and 2 smoothed over $R_s=12h^{-1}$Mpc ({\it Upper
panels}) and $4h^{-1}$Mpc ({\it Lower panels}). We plot the conditional
mean relation $\bar{\delta}_i(\delta_{\rm m})$ from our simulation
results ({\it solid lines}) and from the theoretical prediction of halo
biasing by TS ({\it dashed lines}).  In computing theoretical
predictions, we adjust the range of dark halo mass as our simulated dark
halos (Table~\ref{tab:objects}).

Consider first the results for dark halos. For a given smoothing scale,
the simulated halos exhibit positive biasing for relatively small
$\delta_{\rm m}$ in agreement with the predictions. On the other hand,
they tend to be underpopulated for large $\delta_{\rm m}$, or {\it
anti-biased}.  This is mainly due to the exclusion effect of dark halos
due to their finite volume size as previously discussed in
\citet{Taruya2001} using purely N-body simulations.  The theoretical
model of TS does not take account of this effect, and thus the
discrepancy between the predictions and the simulations becomes more
substantial for smaller $R_{\rm s}$ and/or at lower $z$ as expected.

Since our identified {\it galaxies} have smaller spatial extent than the
halos, the exclusion effect is not so serious. This is clearly
illustrated in lower panels in Figure~\ref{fig:sm412}. In fact they seem
to show much better agreement with the TS predictions despite the fact
that the models are formally valid only for dark halos defined according
to the Press-Schechter manner.

\begin{figure}[tbp]
 \leavevmode
 \begin{center}
  \includegraphics[keepaspectratio,width=15cm]{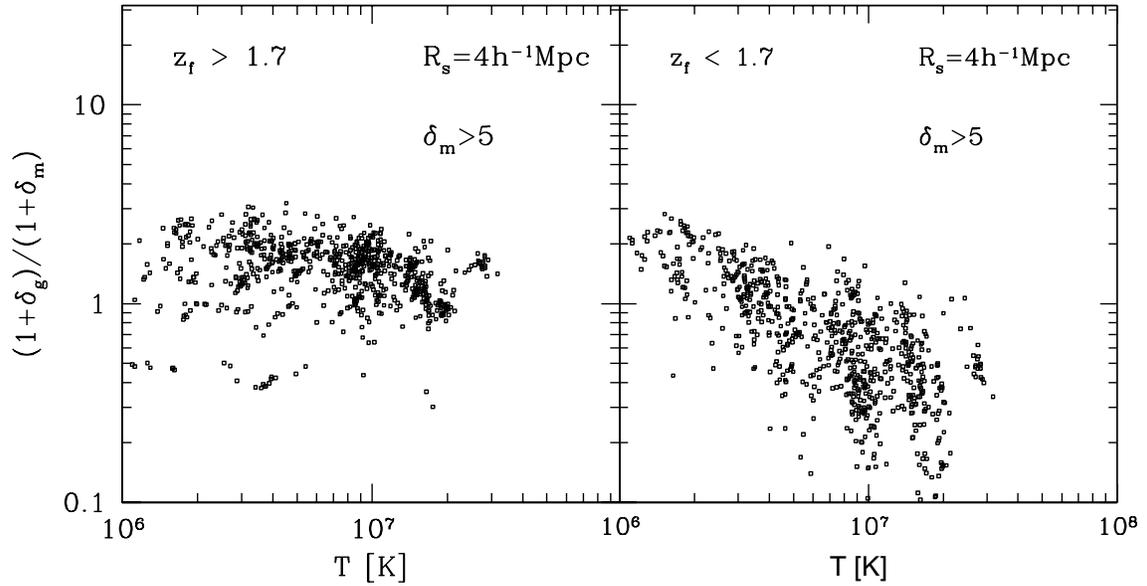}
  \figcaption{Relation between environmental temperature and the ratios
  of dark matter overdensity with that of galaxies with $z_{\rm f}>1.7$
  ({\it left panel}) and $z_{\rm f}<1.7$ ({\it right panel}) in the high
  density regions ($\delta_{\rm m}>5$). Each point corresponds to the
  randomly selected point in the simulation box. The smoothing scale is
  set to $R_s=4h^{-1}$Mpc. \label{fig:gastemp}}
 \end{center}
\end{figure}

A more careful look at the results for galaxies, however, reveals that
$\bar{\delta}_{\rm g}(\delta_{\rm m})/\delta_{\rm m}$ decreases slightly
at larger $\delta_{\rm m}$ especially for smaller smoothing scale
$R_{\rm s}=4h^{-1}$Mpc.  While this tendency may be partially explained
by their volume exclusion effect, their typical sizes seem to be
sufficiently small to account for this.  Rather, we consider two
possible origins of this tendency.  One is the suppression of galaxy
formation at very high temperature and thus high density regions, as
pointed out in \citet{Blanton1999, Blanton2000}.
Figure~\ref{fig:gastemp} shows the dependence of galaxy overdensity on
the surrounding gas temperature separately for galaxies with different
formation redshifts (see \S \ref{sec:zform} for details), and supports
this interpretation; the ratio, $(1+\delta_{\rm g})/(1+\delta_{\rm m})$,
is anti-correlated with the surrounding gas temperature. Comparing the
left and right panels in Figure~\ref{fig:gastemp} indicates that the
anti-correlation with gas temperature is much stronger for galaxies
which form relatively late ($z_{\rm f}<1.7$). On the other hand, those
formed earlier show very weak, at most, anti-correlation, which is
natural because they should have collapsed and formed much before the
surrounding gas acquires the current high temperature.  Another
possibility is that there is an intrinsic difference in formation epoch
of galaxies between over- and under-dense regions. Since, in
hierarchical formation scenarios, objects in over-dense regions tend to
form earlier than those in under-dense regions, it is expected that
young galaxies with $z_{\rm f}<1.7$ form relatively lower-dense thus
low-temperature regions, which is also consistent with
Figure~\ref{fig:gastemp}. The similar analysis for DM cores will
distinguish these two possibilities. Although we notice that the same
correlation exists even for DM cores, we suspect that this is mainly due
to the artificial overmerging effect as we discussed in \S 2.3., and
will revisit this topic with another simulation with higher-resolution
(Yoshikawa, Jing \& Suto, in preparation).

Incidentally, in order to check the dependence of the simulated galaxy
biasing on the lower mass limit of our criteria, $M_{\rm
galaxy}>10^{11}M_{\odot}$ (or equivalently $N_{\rm gas}>40$), we
construct another set of galaxy sample adopting higher mass cutoff
$N_{\rm gas}>80$, and compare their biasing properties.  We find that
the joint probability distribution of $\delta_{\rm m}$ and $\delta_{\rm
g}$ for the galaxy sample selected with $N_{\rm gas}>80$ does not
significantly change from those of the original galaxy sample.

\subsection{Stochasticity and nonlinearity in  biasing of 
galaxies and dark halos}

\begin{figure}[tbp]
 \leavevmode
 \begin{center}
  \includegraphics[keepaspectratio,width=13cm]{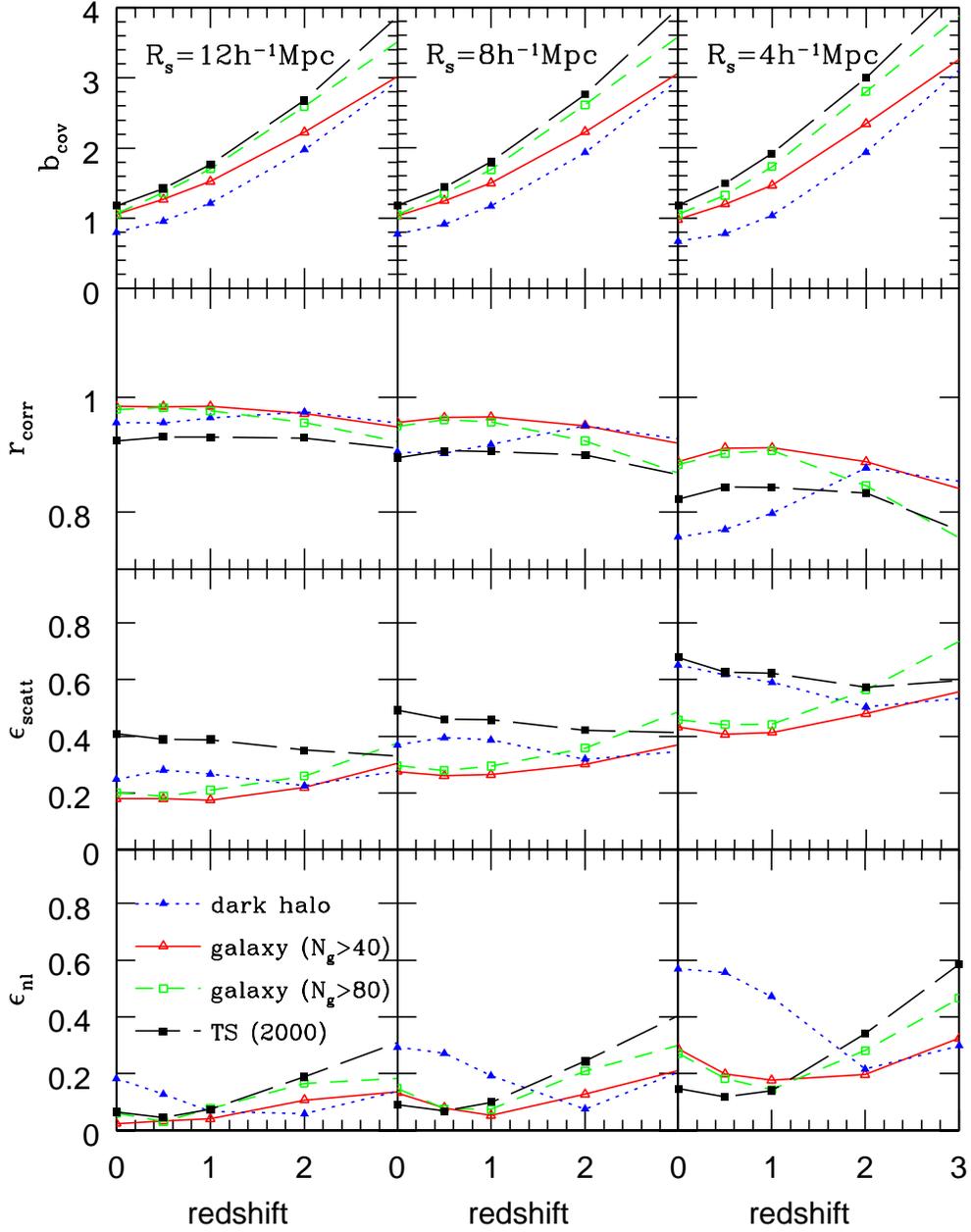}
  \figcaption{Evolution of biasing parameters $b_{\rm cov}$, $r_{\rm
  corr}$, $\epsilon_{\rm scatt}$ and $\epsilon_{\rm nl}$ for galaxies
  (solid lines), dark halos (dotted lines), and the theoretical
  predictions by Taruya \& Suto (2000) (long dashed lines). 
\label{fig:bias_param}}
 \end{center}
\end{figure}

The stochasticity and nonlinearity in galaxy and halo biasing are
clearly identified in Figures~\ref{fig:sm412}. For more quantitative
discussion, we plot in Figure~\ref{fig:bias_param} the evolution of
their biasing parameters $b_{{\rm cov}}$, $r_{{\rm corr}}$,
$\epsilon_{{\rm scatt}}$ and $\epsilon_{{\rm corr}}$ for three different
smoothing radii.

Consider first $b_{{\rm cov}}$. This biasing parameter exhibits strong
time-dependence; the biasing is stronger in the past. This is consistent
with analytic biasing models \citep{Mo1996, Taruya2000}, previous
numerical simulations \citep{Kravtsov1999, Somerville2001, Pearce1999}
and in fact explains the recent observations of Lyman break galaxies
\citep{Giavalisco1998, Adelberger1998} using the halo biasing model
\citep{Mo1996,JS1998}. On the other hand, the scale-dependence of
$b_{{\rm cov}}$ is very weak as in the case of the biasing parameter
defined through the two-point correlation function (see the next
subsection).

Both $\epsilon_{\rm scatt}$ and $\epsilon_{\rm nl}$ in our simulated
catalogues are somewhat smaller than the TS prediction, but their
qualitative behavior is consistent with the model; larger on small
scales and almost independent of $z$. The biasing becomes linear and
deterministic relation and also the volume exclusion is less effective
for larger smoothing scales.  The current degree of the stochasticity
and nonlinearity does not hardly affect the amplitude of clustering (see
eq.[\ref{eq:bvar}]), but the topology of the isodensity contours is
sensitive to the nonlinearity even at this level \citep{Hikage2001}.

It is interesting to notice that the biasing parameters for galaxies
show similar behavior and are closer to the predicted behavior. In
addition, all biasing parameters for dark halos and galaxies behave very
similarly at high redshifts $z\simeq 2-3$. This indicates that the
spatial distribution of galaxies and dark halos are statistically
similar, and can be understood by the fact that we have one-to-one
correspondence between dark halos and galaxies at $z\simeq 2-3$ as shown
below.

Figure~\ref{fig:bias_param} also shows that the evolution of biasing is
almost independent on the lower mass limit of the galaxies. This might
be interpreted as our simulated galaxy sample is nearly complete for the
present purpose.

\begin{figure}[tbp]
 \leavevmode
 \begin{center}
  \includegraphics[keepaspectratio,width=16cm]{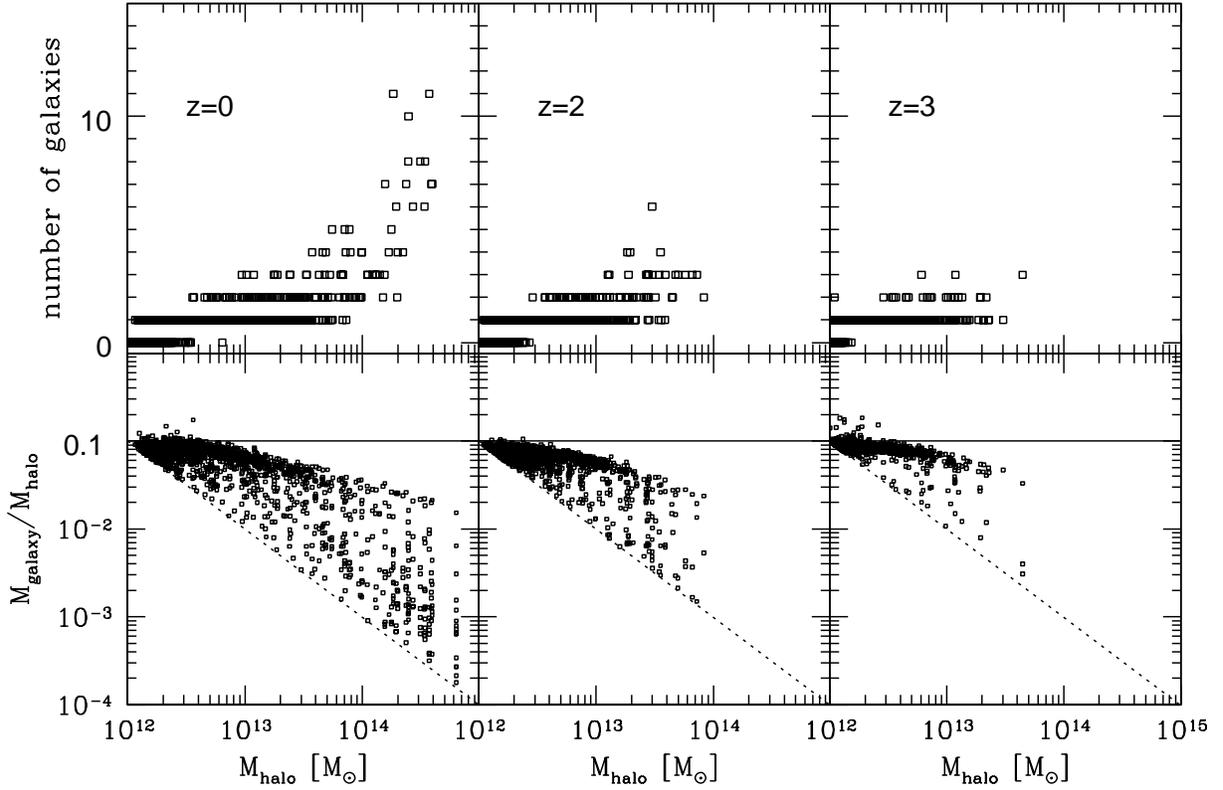}
  \figcaption{Number of galaxies which reside in a dark halo ({\it upper
  panels}) and distributions of mass ratio between galaxies and dark
  halos which host them ({\it lower panels}) at redshift $z=0$ and
  2. Solid and dashed lines in lower panels indicate the mean baryon
  fraction ($\Omega_{\rm b}/\Omega_0$) and resolution limit constrained
  by the minimum mass of galaxies in their definition (see
  \S\ref{sec:galdef}), respectively.  \label{fig:massratio}}
 \end{center}
\end{figure}

Figure~\ref{fig:massratio} shows the number of member galaxies which
reside within the virial radius of their hosting dark halos ({\it upper
panels}) and the distribution of their mass ratios ({\it lower panels})
at redshift $z=0$, 2 and 3. Solid and dashed lines in lower panels
indicate the cosmic mean baryon fraction $\Omega_{\rm b}/\Omega_0$ and
resolution limit of galaxy mass ($M_{\rm galaxy}=10^{11}\,M_{\odot}$),
respectively. One can see that most dark halos at $z=3$ host only one
galaxy, explicitly justifying the {\it empirical} assumption of
one-to-one correspondence between dark halos and Lyman-break galaxies
around $z=3$ in previous studies \citep{JS1998, Steidel1998}. The
subsequent evolution of dark halos involves several merger processes,
and thus dark halos at lower redshifts tend to host multiple member
galaxies.

\subsection{Biasing in terms of the two-point correlation function}

The previous subsections discuss only the biasing parameters defined
from the one-point statistics.  In this subsection, we turn to a more
conventional biasing parameter defined through the two-point statistics:
\begin{equation}
 b_{\xi,i}(r)\equiv \sqrt{\frac{\xi_{ii}(r)}{\xi_{\rm mm}(r)}} ,
\end{equation}
where $\xi_{ii}(r)$ and $\xi_{\rm mm}(r)$ are two-point correlation
functions of objects $i$ and of dark matter, respectively.  While the
above biasing parameter is ill-defined where either $\xi_{ii}(r)$ or
$\xi_{\rm mm}(r)$ becomes negative, it is not the case at clustering
scales of interest ($< 10 h^{-1}$Mpc).  The relation of one-point and
two-point biasing parameters is also investigated in detail by
\citet{Taruya2001} for density peaks and dark halos.

\begin{figure}[tbp]
 \leavevmode
 \begin{center}
  \includegraphics[keepaspectratio,width=14cm]{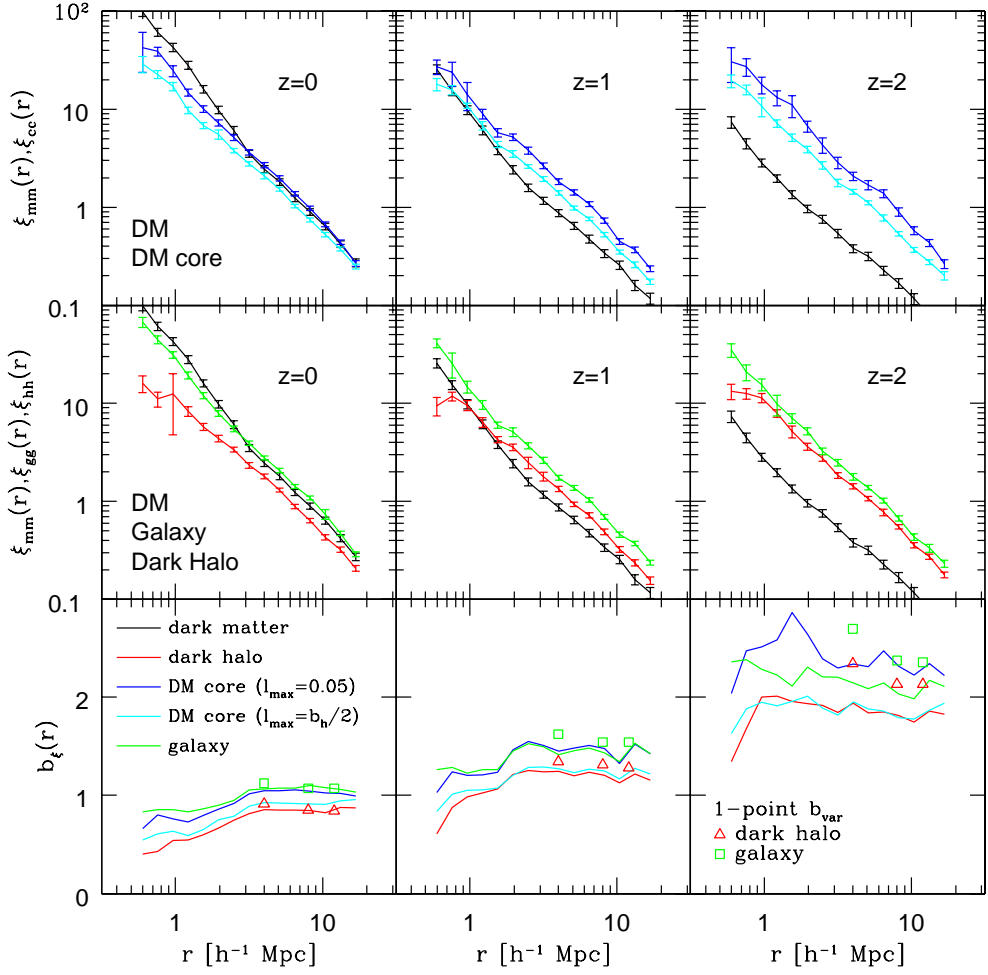}
  \figcaption{
  Upper panels show two-point correlation functions for dark matter and
  DM cores at redshift $z=0$, 1 and 2. Middle panels for those of dark
  matter, galaxies and dark halos. In lower panels, the profiles of
  biasing parameter $b_{\xi}(r)$ for dark halos, DM cores and galaxies
  at $z=0.0$, 1 and 2 are shown. In lower panels, we also plot the
  parameter $b_{\rm var}$ on the smoothing scale $R_s=4h^{-1}$Mpc,
  $8h^{-1}$Mpc and $12h^{-1}$Mpc at $r=R_s$ for each kind of objects by
  different symbols.\label{fig:xi_bias}}
 \end{center}
\end{figure}

Figure~\ref{fig:xi_bias} shows two-point correlation functions of dark
matter, galaxies, dark halos and DM cores ({\it upper and middle
panels}), and the profiles of biasing parameters $b_{\xi}(r)$ for those
objects ({\it lower panels}) at $z=0$, 1 and 2. In the upper panels, we
show the correlation functions of DM cores identified with two different
maximum linking length; $l_{\rm max}=0.05$ as presented in
\S~\ref{subsec:DMcore} and $l_{\rm max}=b_{\rm h}/2$. Correlation
functions of DM cores identified with $l_{\rm max}=0.05$ are similar to
those of galaxies. On the other hand, those identified with $l_{\rm
max}=b_{\rm h}/2$ exhibit much weaker correlation, and are rather
similar to those of dark halos. This is due to the fact that HFOF
algorithm with larger $l_{\rm max}$ tends to pick up lower mass halos
which are poorly resolved in our numerical resolution.

The correlation functions of galaxies are almost unchanged with
redshift, and that of dark halos only slightly evolves between $z=0$ and
2. By contrast, the amplitude of the dark matter correlation function
evolves rapidly by factor of $\sim 10$ from $z=2$ to $z=0$. The biasing
parameter $b_{\xi,{\rm g}}$ is larger at a higher redshift, for example,
$b_{\xi,{\rm g}}\simeq 2$--2.5 at $z=2$. These results are consistent
with the numerical studies by \citet{Bagla1998}, \citet{Colin1999} and
\citet{Pearce1999} and also qualitatively explains the clustering of
Lyman-break galaxies \citep{Giavalisco1998}. The biasing parameter
$b_{\xi,{\rm h}}$ for dark halos is systematically lower than that of
galaxies and DM cores again due to the volume exclusion effect. At
$z=0$, galaxies and DM cores are slightly anti-biased relative to dark
matter at $r\simeq 1h^{-1}$Mpc, which is also consistent with previous
numerical simulations \citep{Pearce1999, Colin1999, Benson2000,
Somerville2001} and also with the observational results from the Las
Campanas Redshift Survey \citep{JMB1998}. In lower panels, we also plot
the one-point biasing parameter $b_{{\rm var},i} \equiv
\sigma_i/\sigma_{\rm m}$ at $r=R_s$ for comparison. In general we find
that $b_{\xi,i}$ is very close to $b_{{\rm var},i}$ at $z \sim 0$, but
systematically lower than $b_{{\rm var},i}$ at higher redshifts.

\section{THE FORMATION EPOCH AS AN ORIGIN OF THE MORPHOLOGICAL TYPE OF
 GALAXIES\label{sec:zform}}

It is fairly established that there exists a certain correlation between
the morphology of galaxies and their star formation history; early-type
galaxies form via initial star bursts at high redshifts while late-type
galaxies experience continuous and relatively mild star formation
history \citep{Roberts1994,Kennicutt1998}. This implies that the galaxy
morphology is empirically related to its formation epoch.  On the basis
of this interpretation, one can examine the morphology-dependent
clustering of galaxies by classifying our simulated galaxies according
to their formation epoch.

We have fifty outputs of all simulation particles at different redshifts
between $z=9$ and $0$.  For each galaxy identified at $z=0$, we define
its formation redshift $z_{\rm f}$ by the epoch when half of its {\it
cooled gas} particles satisfy the criteria (\ref{eq:decouple}) and
(\ref{eq:densthresh}). Roughly speaking, $z_{\rm f}$ corresponds to the
median formation redshift of {\it stars} in the present-day galaxies. We
divide all simulated galaxies at $z=0$ into two populations (the young
population with $z_{\rm f}<1.7$ and the old population with $z_{\rm
f}>1.7$) so as to approximate the observed number ratio of $3/1$ for
late-type and early-type galaxies \citep{Loveday1995}.

\begin{figure}[tbp]
 \leavevmode
 \begin{center}
  \includegraphics[keepaspectratio,width=14cm]{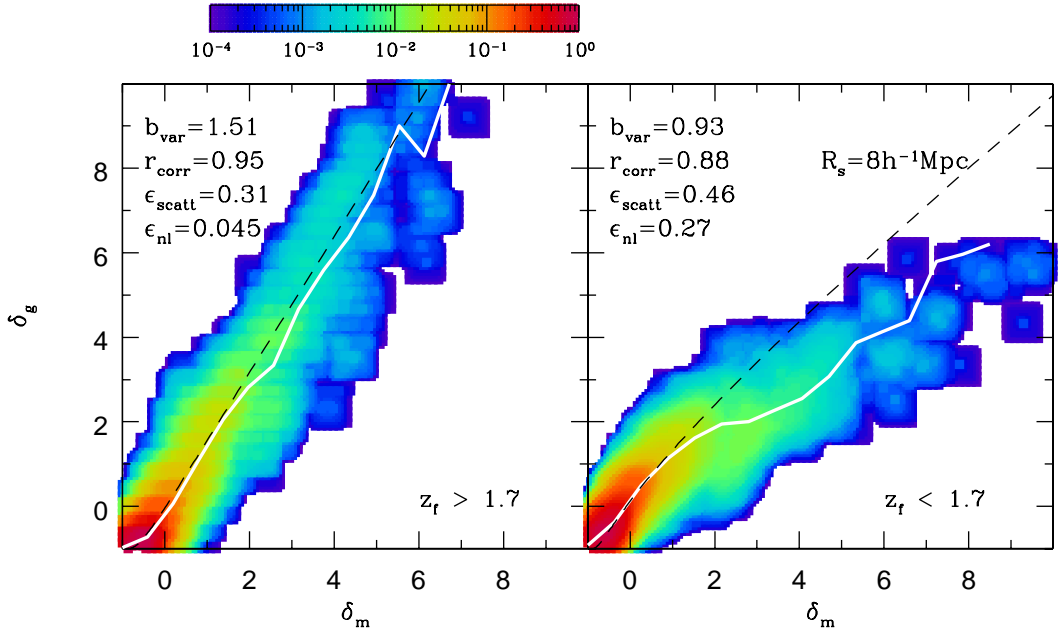}
  \figcaption{Joint probability distributions of density fields of dark
  matter and galaxies with different formation epochs on the scale of
  $R_s=8h^{-1}$Mpc. Left panel is for galaxies with $z_{\rm f}>1.7$ and
  right panel for ones with $z_{\rm f}<1.7$. Solid lines indicate the
  simulated mean relations. For comparison, the predictions of mean
  biasing for dark halos with their formation redshift greater and less
  than 1.7. are shown in left and right panels,
  respectively. \label{fig:early_late_bias}}
 \end{center}
\end{figure}

Figure~\ref{fig:early_late_bias} shows the joint probability
distribution of $\delta_{\rm m}$ and $\delta_{\rm g}$ respectively for
the old ({\it left panel}) and young ({\it right panel})
populations. They exhibit clear difference in their clustering
properties. Their biasing parameters are $\sigma_{\rm g}=1.73 (1.06)$,
$b_{\rm var,g}=1.51 (0.93)$ and $r_{\rm corr,g}=0.95 (0.88)$ for the old
(young) population. These results qualitatively agree with
\citet{Blanton1999}, and \citet{Somerville2001} also showed a similar
result that red galaxies are biased compared to the overall population
and blue ones are anti-biased, where galaxies with color $B-V>0.8$ are
defined as red galaxies and the remainder as blue ones. The dashed lines
in both panels indicate the TS predictions of the mean biasing for dark
halos restricting the formation epoch as $z_{\rm f}>1.7$ and $z_{\rm
f}<1.7$, respectively; the old population shows excellent agreement with
the halo biasing prediction while the young population behaves rather
differently.  This indicates that early-type galaxies preferentially
reside in the center of the massive halos almost in a one-to-one manner
and that late-type galaxies avoid the dense environment, which is
consistent with the observed morphology-density relation
\citep{Dressler1980,Postman1984,Dressler1997}.

\begin{figure}[tbp]
 \leavevmode
 \begin{center}
  \includegraphics[keepaspectratio,width=14cm]{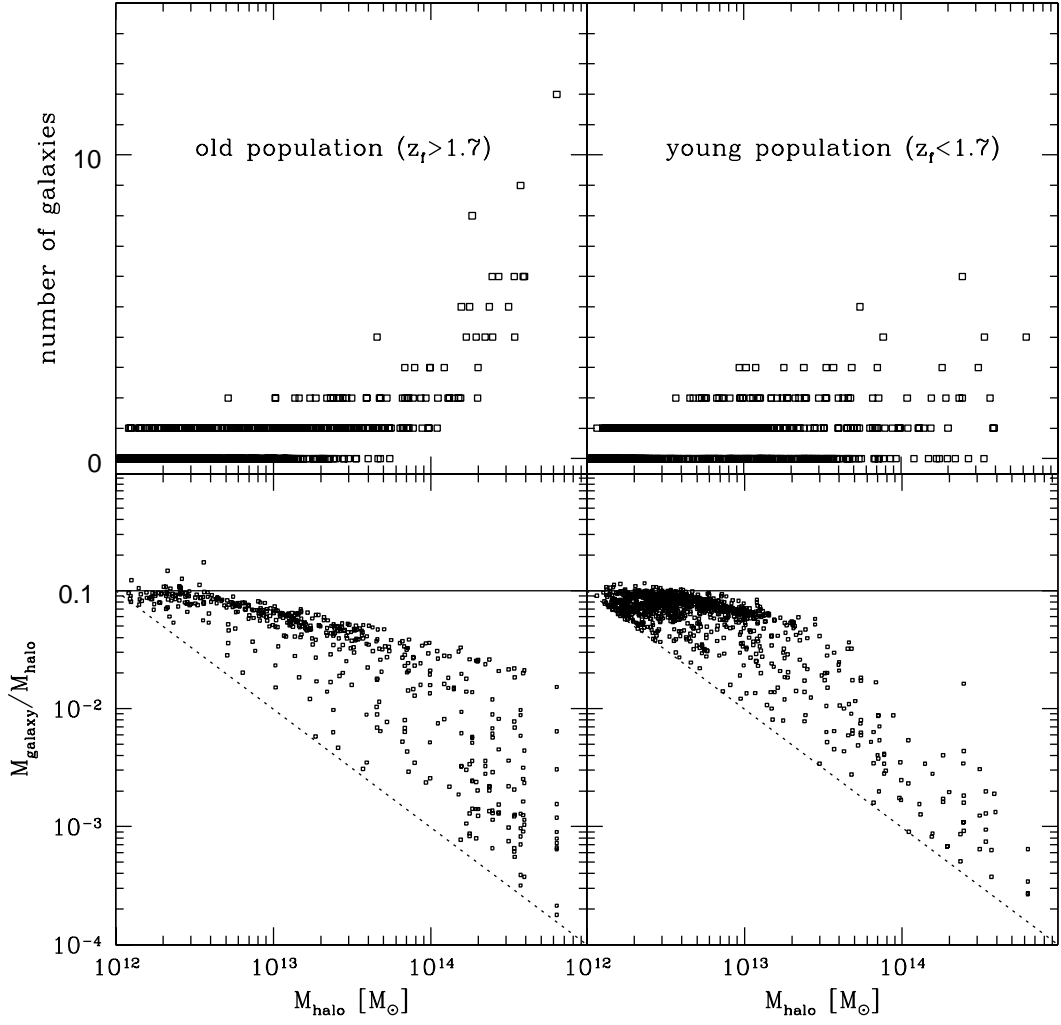}
  \figcaption{Same as Figure~\ref{fig:massratio} except for old ({\it
  left}) and young ({\it right}) populations of galaxies at
  $z=0$. \label{fig:early_late_massratio}}
 \end{center}
\end{figure}

This interpretation is directly confirmed in
Figure~\ref{fig:early_late_massratio}.  Massive halos have a larger
fraction of the old population of galaxies, while the young population
of galaxies mainly reside in smaller halos.  This segregation may be
understood by the same mechanisms of anti-bias of galaxies at high
density regions. As discussed in \S~\ref{subsec:bias}, due to the
suppression of galaxy formation in high temperature regions at lower
redshift, and/or a different formation epoch for over- and under-dense
regions, we have the deficiency of the young population of galaxies
within massive dark halos at $z=0$, and galaxies formed at high redshift
trapped within the gravitational potential of dark halos gradually tend
to trace the distribution of underlying dark matter.

\begin{figure}[tbp]
 \leavevmode
 \begin{center}
  \includegraphics[keepaspectratio,width=14cm]{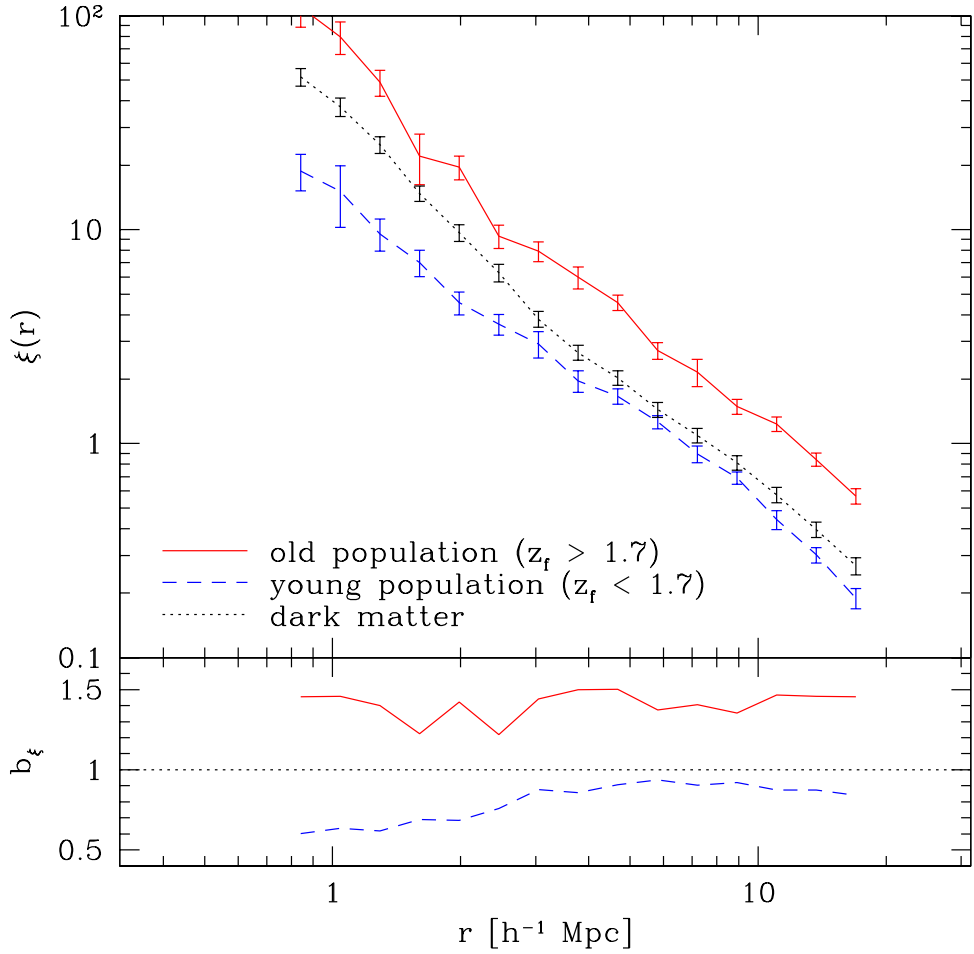}
  \figcaption{Two-point correlation functions for the old and young
  populations of galaxies at $z=0$ as well as that of dark matter
  distribution. The profiles of bias parameters $b_{\xi}(r)$ for both of
  the two populations are also shown in the lower panel.
  \label{fig:early_late_xi_bias}}
 \end{center}
\end{figure}

The difference of the clustering amplitude can be also quantified by
their two-point correlation functions at $z=0$ as plotted in
Figure~\ref{fig:early_late_xi_bias}.  The old population indeed clusters
more strongly than the mass, and the young population is
anti-biased. The relative bias between the two populations $b^{\rm
rel}_{\xi,{\rm g}} \equiv \sqrt{\xi_{\rm old}/\xi_{\rm young}}$ ranges
$1.5$ and 2 for $1h^{-1}\mbox{Mpc}<r<20h^{-1}\mbox{Mpc}$, where
$\xi_{\rm young}$ and $\xi_{\rm old}$ are the two-point correlation
functions of the young and old populations.  Again this is completely
consistent with the observational indications that the clustering of
early-type galaxies is stronger than that of late-type galaxies by a
factor of 3--4 in terms of the amplitude of two-point correlation
functions \citep{Loveday1995, Hermit1996}.

All the above results suggest that the old and young populations of
galaxies in our simulations may be interpreted as the early-type and
late-type galaxies in the present universe, and that the formation epoch
and the hydrodynamical environment play the important role in
determining the morphology of galaxies.  We note here that the above
result is fully consistent with the recent analysis of the IRAS PSCz
galaxy sample by \citet{JBS2001}, who found a strong anti-bias of the
IRAS-selected galaxies (and thus mainly late-types). The degree of the
detected bias is accounted for by the phenomenological
cluster-underweight bias model \citep{JMB1998}, and also by the
semi-analytic modeling of galaxy formation which assumes that the galaxy
morphology is determined by the frequency of the major merger of halos
\citep{Kauffmann1996, Kauffmann1997, Kauffmann1999}.

\section{CONCLUSIONS AND DISCUSSION}

Using a cosmological SPH simulation, we directly simulate the formation
of galaxies via radiative cooling of baryonic component and identify
galaxies as isolated and distinct groups of cold gas particles.  We
calculated the biasing of galaxies and dark halos, and in particular,
compared their properties with the theoretical prediction of the halo
biasing model proposed by TS.

Our major findings are summarized as follows;

(1) The clustering of dark halos suffers from the the volume exclusion
effect due to their finite size, especially at small scales.  On the
other hand, the halo biasing model by TS can reasonably account for the
clustering of ``galaxies'' at large scales. At smaller scales, however,
galaxies are anti-biased relative to dark matter at high density and
thus high temperature environment.

(2) The biasing parameters are strongly time-dependent.  At $z\sim 3$,
our galaxies exhibit strong biasing; $b_{{\rm cov,g}}\simeq 3\mbox{--}4$
and $b_{\xi,{\rm g}}\simeq 3$, which is consistent with the observed
clustering of Lyman-break galaxies \citep{Adelberger1998,
Giavalisco1998, Steidel1998}.

(3) The formation epoch $z_{\rm f}$ is the major parameter in
determining the morphological type of galaxies. In our specific example,
galaxies identified at $z=0$ with $z_{\rm f}>1.7$ and $z_{\rm f}<1.7$
can be roughly regarded as early-type and late-type galaxies,
respectively. The former tightly correlates with the massive host halos
and shows stronger clustering, while the latter is anti-biased and more
stochastic.  These suggest that biasing properties of galaxies,
identified by different photometric bands or color selections, should be
significantly different, which should be kept in mind in comparing the
galaxy clustering from different galaxy catalogues.

Our current definition of galaxies in simulation data is admittedly
rather phenomenological. Apparently more observationally oriented
classification of galaxies, for example using color or magnitude of
galaxies, is necessary for direct comparison with observations. We plan
to implement more realistic prescriptions of galaxy formation and
evolution including star formation, feedback and UV background heating
in due course.  Nevertheless it is quite encouraging that even a simple
scheme described here explains the major properties of galaxy clustering
in the universe.

\bigskip

We thank an anonymous referee for many useful comments, in particular
for important comments on the dependence of galaxy biasing on their
formation epoch and the importance of the artificial overmerging effect
in the current simulations.  We also thank S. Gottl\"{o}ber for a
careful reading of the manuscript. K.Y. and A.T. gratefully acknowledge
support from JSPS (Japan Society for the Promotion of Science)
fellowships. Y.P.J. is supported in part by the One-Hundred-Talent
Program, by NKBRSF (G19990754) and by NSFC.  Numerical computations were
carried out at ADAC (the Astronomical Data Analysis Center) of the
National Astronomical Observatory, Japan and at KEK (High Energy
Accelerator Research Organization, Japan). This research was supported
in part by the Grant-in-Aid by the Ministry of Education, Science,
Sports and Culture of Japan (07CE2002, 12640231), and by the
Supercomputer Project (No.99-52, No.00-63) of KEK.

\begin{deluxetable}{cccc}
 \footnotesize 
 
 \tablecaption{Number and mass range of identified objects and adopted
 linking length for FOF algorithm. \label{tab:objects}} 
 \tablewidth{0pt}

 \tablehead{\colhead{redshift} & \colhead{dark halo} & \colhead{galaxy}
 & \colhead{DM core} }
 \startdata
 0.0 & 1797 ($0.164\,\bar{l}$) & 1604 ($0.0164\,\bar{l}$)& 1525
 ($0.05\,\bar{l}$) \\
 & $10^{12}M_\odot \sim 8.6\times10^{14}M_\odot$ 
 & $10^{11}M_\odot \sim 9.5\times10^{12}M_\odot$ 
 & $4.3\times 10^{11}M_\odot \sim 2.0\times10^{14}M_\odot$\\ 
 0.5 & 2105 ($0.184\,\bar{l}$) & 1936 ($0.0246\,\bar{l}$)& 1721 
 ($0.05\,\bar{l}$)\\
 & $10^{12}M_\odot \sim 3.5\times10^{14}M_\odot$ 
 & $10^{11}M_\odot \sim 6.8\times10^{12}M_\odot$ 
 & $4.3\times 10^{11}M_\odot \sim 6.3\times10^{13}M_\odot$ \\
 1.0 & 2201 ($0.192\,\bar{l}$) & 1861 ($0.0328\,\bar{l}$)& 1543 
 ($0.05\,\bar{l}$)\\
 & $10^{12}M_\odot \sim 2.3\times10^{14}M_\odot$ 
 & $10^{11}M_\odot \sim 3.6\times10^{12}M_\odot$ 
 & $4.3\times 10^{11}M_\odot \sim 2.9\times10^{13}M_\odot$ \\
 2.0 & 1859 ($0.197\,\bar{l}$) & 1360 ($0.0492\,\bar{l}$)&  765 
 ($0.05\,\bar{l}$)\\
 & $10^{12}M_\odot \sim 8.2\times10^{13}M_\odot$ 
 & $10^{11}M_\odot \sim 2.0\times10^{12}M_\odot$ 
 & $4.3\times 10^{11}M_\odot \sim 9.8\times10^{12}M_\odot$ \\
 3.0 & 1165 ($0.199\,\bar{l}$) &  996 ($0.0656\,\bar{l}$)&  278 
 ($0.05\,\bar{l}$)\\
 & $10^{12}M_\odot \sim 3.8\times10^{13}M_\odot$ 
 & $10^{11}M_\odot \sim 1.4\times10^{12}M_\odot$ 
 & $4.3\times 10^{11}M_\odot \sim 4.2\times10^{12}M_\odot$ \\
 \enddata
\end{deluxetable}

\clearpage

\clearpage

\begin{figure}[tbp]
 \leavevmode
 \begin{center}
  \figcaption{Distribution of gas particles, dark matter particles,
  galaxies and dark halos in the volume of $75h^{-1}\times
  75h^{-1}\times 30h^{-1}$Mpc$^3$ model at $z=0$. {\it Upper-right:}gas
  particles; {\it Upper-left:} dark matter particles; {\it Lower-right:}
  galaxies; {\it Lower-left:} DM cores\label{fig:LCDM_Z00}}
 \end{center}
\end{figure}
\begin{figure}[tbp]
 \leavevmode
 \begin{center}
  \figcaption{Same as Figure~\ref{fig:LCDM_Z00} but for $z=2$.
  \label{fig:LCDM_Z20}}
 \end{center}
\end{figure}
\begin{figure}[tbp]
 \leavevmode
 \begin{center}
  \figcaption{Snapshots of the most massive cluster ($M \simeq 8\times
  10^{14}M_{\odot}$) in the simulation at $z=0$. {\it Upperleft:} dark
  matter; {\it Upper-right:} gas; {\it Lower-left:} DM cores; {\it
  Lower-right:} cold gas. Circles in lower panels indicate the positions
  of galaxies identified according to our criteria. The comoving size of
  the box is $6.25h^{-1}$Mpc per side.
  \label{fig:region_1}}
 \end{center}
\end{figure}
\begin{figure}[tbp]
 \leavevmode
 \begin{center}
  \figcaption{Same as Figure~\ref{fig:region_1} but for a poorer cluster
  with $M \simeq 10^{14}M_{\odot}$.
  \label{fig:region_2}}
 \end{center}
\end{figure}
\begin{figure}[tbp]
 \leavevmode
 \begin{center}
  \figcaption{Joint probability distributions of overdensity fields for
  dark halos and galaxies with dark matter overdensity smoothed over
  $R_s=12h^{-1}$Mpc ({\it Upper panels}) and $R_s=4h^{-1}$Mpc ({\it
  Lower panels})
  at redshift $z=0$, 1 and 2. Solid lines indicate the
  conditional mean $\bar{\delta}_i(\delta_{\rm m})$ for each
  object. Dashed lines in each panel depict the theoretical prediction
  of conditional mean by Taruya \& Suto (2000). 
  \label{fig:sm412}}
 \end{center}
\end{figure}
\clearpage
\clearpage
\end{document}